\documentstyle[preprint,aps,prd,epsfig]{revtex}
\begin{document}  
\draft
\preprint{OCIP/C 95-11, hep-ph/9508264}
\title{THE $\xi(2220)$ REVISITED: \\
STRONG DECAYS OF THE $1^3F_2$ and $1^3F_4$ $s\bar{s}$ MESONS}
\author{Harry G. Blundell\footnote{e-mail:  harry@physics.carleton.ca} 
and Stephen Godfrey\footnote{e-mail: godfrey@physics.carleton.ca}}
\address{Ottawa-Carleton Institute for Physics \\
Department of Physics, Carleton University, Ottawa CANADA, K1S 5B6}
\date{July 1995}
\maketitle
\begin{abstract}
We calculated the decay widths of the $1^3F_2$ and $1^3F_4$ 
$s\bar{s}$ mesons and compared them to 
the measured properties of the $\xi(2220)$ (now known
as the $f_4(2220)$).  Including previously neglected decay modes we found
that the width of the $^3F_2$ state $s\bar{s}$ meson is much larger than
previously believed making this explanation unlikely.  
On the other hand the predicted width  of the $^3F_4$ state, 
although broader than the observed width, is consistent within the 
uncertainties of the model.  This interpretation predicts large 
partial widths to $K K^*(892)$ and $K^*(892) K^*(892)$ final states which 
should be 
looked for.  A second possibility that would account for the 
different properties of the $\xi(2220)$ seen in different experiments 
is that two hadronic states exist at this mass.  The first would be a broader 
$^3F_4$ $s\bar{s}$ state which is seen in hadron beam experiments 
while the second would be a narrow state with high glue content seen in 
the gluon rich $J/\psi$ radiative decay.  Further experimental 
results are needed to sort this out.
\end{abstract}

\section{INTRODUCTION}
\label{sec:intro}

It is roughly a decade since the $\xi(2220)$, now known as the 
$f_4(2220)$, was discovered by the MARK III collaboration
in $J/\psi$ radiative decays to 
$K^+ K^-$ and $K_S K_S$ 
final states \cite{markiii}.  Its most interesting property, which 
attracted considerable attention, was its narrow width of roughly 
30~MeV.  Because the width was inconsistent with expectations for a 
conventional $q\bar{q}$ meson with such a large mass, the $\xi$'s 
discovery led to speculation that it might be
a Higgs boson \cite{higgs}, 
a bound state of coloured scalars \cite{scalar}, 
a four quark state \cite{4quark,pakvasa}, 
a $\Lambda \bar{\Lambda}$ bound state \cite{ono87},
a hybrid \cite{hybrid}, 
or a glueball \cite{glueball}.
Despite the prevailing wisdom, the authors of 
Ref.~\cite{godfrey84,pakvasa}
argued that the properties of the $\xi(2220)$ could be consistent with 
those of a conventional meson:  the L=3 $s\bar{s}$ meson with 
$J^{PC}=2^{++}$ or $J^{PC}=4^{++}$.  

In the original analysis of L=3 $s\bar{s}$ properties it was shown 
that of the $q\bar{q}$ states with the appropriate $J^{PC}$ 
quantum numbers only the 
$^3F_2$ and $^3F_4$ $s\bar{s}$ states 
of the first L=3 multiplet have masses 
consistent with the $\xi(2220)$ \cite{godfrey84}.  
According to this analysis these 
two states were exceptional in that they have a limited number of 
available decay modes which are all relatively weak.  However, the
analysis was not exhaustive in that it did not calculate the decay 
widths to all possible final states.  In particular it made the 
assumption, which we will see to be incorrect, that the decays to 
an $L=1$ meson and a $K$ or $\eta$ were small on the basis of phase space 
arguments alone.  

To further complicate the discussion, more recent experiments have 
observed a hadronic state decaying to $K\bar{K}$ in different reactions and 
with different properties. 
The various experimental results relevant to the $\xi(2220)$ are 
summarized in Table \ref{table1}.
The most recent measurement of the 
$\xi(2220)$ properties by the BES collaboration \cite{bes}
indicates that its decays are approximately 
flavour symmetric giving support to the glueball interpretation.
At the same time, although the narrow 
$\xi(2220)$ was not seen in $J/\psi$ radiative decays  by the DM2 experiment 
despite the fact that DM2 has slightly higher statistics, DM2 did 
observe a broader state decaying into $K\bar{K}$ \cite{dm2}.  If all 
the experiments are taken at face value the overall picture is 
confused and contradictory.

In this paper we re-examine the nature of the $\xi(2220)/f_4(2220)$ meson 
and calculate the partial widths of the $^3F_2$ and $^3F_4$ 
$s\bar{s}$ states to all OZI-allowed 2-body final states allowed 
by phase space.  To give a measure of the reliability of our 
analysis we calculate the widths using both the $^3P_0$ decay model (often 
referred to as the quark-pair creation 
decay model) \cite{leyaouanc73,roberts} 
and the flux-tube breaking decay model \cite{kokoski87}.  As an additional 
consistency check we 
calculated several partial widths using the pseudoscalar decay 
model \cite{godfrey85}.  
Our goal is to shed some light on the nature of the 
$\xi(2220)$ by comparing the quark model predictions for the hadronic 
widths to the various experimental results.

The outline of the paper is as follows.  In section \ref{sec2} we briefly 
outline the models of hadron decays and the
fitting of the parameters of the models. We relegate the details to 
the appendices.  In section \ref{sec3} we present 
the results of our calculations for the $L=3$ mesons and discuss our 
results.  In the final section we attempt to make sense of the 
various contradictory experimental results and put forward our 
interpretation along with some suggested measurements which may 
clear up the situation.

\section{MODELS OF MESON PROPERTIES AND DECAYS}
\label{sec2}

The quark model has proven to be a useful tool to describe the 
properties of hadrons. The quark model
has successfully described weak, electromagnetic, and strong 
couplings \footnote{See for example Ref.  \cite{godfrey85}.}.
In some cases we will use simplified meson wavefunctions which have
been used elsewhere to describe hadronic decays \cite{kokoski87} 
while in other cases we will 
use more complicated wavefunctions from a relativized quark model 
which includes one-gluon exchange 
and a linear confining potential \cite{godfrey85}.  
The strong decay analysis was performed using the QCD based
flux-tube breaking
model \cite{kokoski87}.  It has the attractive feature of
describing decay rates to all possible 
final states in terms of just one fitted parameter.   
We also include results for the $^3P_0$ model, often referred to 
as the quark-pair creation model \cite{leyaouanc73,roberts}, which is a 
limiting case of 
the flux-tube breaking model and which greatly simplifies the calculations 
and gives similar results.  As a final check we calculated 
some partial widths using the pseudoscalar emission model \cite{godfrey85}
and confirmed that it also gave  results similar to those
of the flux-tube breaking model.

\subsection{Decays by the $^3P_0$ Model}

The $^3P_0$ model \cite{leyaouanc73,roberts} is applicable to OZI-allowed
strong decays of a meson into two other mesons, as well as the two-body strong
decays of baryons and other hadrons.
Meson decay occurs when a
quark-antiquark pair is produced from the vacuum in a state suitable for
quark rearrangement to occur, as in Fig.~\ref{3p0decay}.
The created pair will have the quantum numbers of the vacuum, $^3P_0$.
There is one undetermined parameter $\gamma$ in the model - it represents
the probability that a quark-antiquark pair will be created from the vacuum.
The rest of the model is just the description of the overlap of the initial
meson (A) and the created pair with the two final mesons (B,C), to calculate
the probability that rearrangement (and hence decay) will occur.  
A brief description of the model is included in Appendix \ref{appa}, and the 
techniques by which the calculations were performed are discussed in Appendices
\ref{appc} and \ref{appd}.

\subsection{Decays by the Flux-Tube Breaking Model}

In the flux-tube picture a meson consists of a quark and antiquark connected by
a tube of chromoelectric flux, which is treated as a vibrating string.  For
mesons the string is in its vibrational ground state.  Vibrational excitations
of the string would correspond to a type of meson hybrid, particles whose
existence have not yet been confirmed.

The flux-tube breaking decay model \cite{kokoski87} is similar to the
$^3P_0$ model, but extends it by considering the actual dynamics of the
flux-tubes.  This is done by including a factor representing the overlap of 
the flux-tube of the initial meson with those of the two outgoing mesons.  
A brief review of the model is given in Appendix \ref{appb}, and the 
techniques by which the calculations were performed are discussed in Appendices
\ref{appc} and \ref{appd}.

\subsection{Fitting the Parameters of the Decay Models}

The point of these calculations is to obtain a reliable estimate 
of the $^3F_2$ and $^3F_4$ $s\bar{s}$ meson decay 
widths.  To do so we considered several variations of the 
flux-tube breaking model.  By seeing how much the results vary under the 
various assumptions we can estimate the reliability of the predictions.

The first variation lies with the normalization of the mock meson wavefunctions
and the phase space used to calculate the decay widths
\cite{geiger94}.  In the Appendices we
have normalized the mock meson wavefunctions relativistically to $2 E$ and used
relativistic phase space, which leads to a factor of $E_B E_C/M_A$ in the final
expression for the width in the centre of mass frame.  We will refer to this as
relativistic phase space/normalization (RPSN).  However, there are arguments
\cite{isgurpc} that heavy quark effective theory fixes the assumptions in the
mock meson prescription and suggests that the energy factor be replaced by
$\widetilde{M}_B \widetilde{M}_C/\widetilde{M}_A$, where the $\widetilde{M}_i$
are the calculated masses of the meson $i$ in a spin-independent
quark-antiquark potential \cite{kokoski87}.  (In other words $\widetilde{M}_i$
is given by the hyperfine averaged mass that is equal to the centre of gravity
of the triplet and singlet masses of a multiplet of given $L$.)  We will refer
to this as the Kokoski-Isgur phase space/normalization (KIPSN).

The second variation in our 
results is the choice of wavefunctions.  We calculate decay widths 
for two cases.  In the first we use simple harmonic oscillator (SHO) 
wavefunctions with a common oscillator parameter for all mesons.  
In the second case we use the 
wavefunctions, calculated in a relativized quark model, of 
Ref. \cite{godfrey85} which we will label RQM.  
In all we looked at six 
cases:  the $^3P_0$ model using the SHO wavefunctions, 
the flux-tube breaking model again using the SHO wavefunctions,
and the flux-tube breaking model using 
the RQM wavefunctions of Ref. \cite{godfrey85}; in all three cases 
we used both choices of phase space/normalization.

Some comments about the details of the calculations are in order.  
For the SHO
wavefunctions, we took for the oscillator parameter $\beta = 400$~MeV
which is the value used by Kokoski and Isgur \cite{kokoski87}.  
However,  different quark models find different values of $\beta$ so 
that there is the question of the sensitivity of our results to 
$\beta$.  We will address this issue below.
We used quark masses in the ratio $m_u:m_d:m_s = 3:3:5$ --- this
differs from the calculations of Ref. \cite{kokoski87}, which
ignored the strange-quark mass difference. In the RQM wavefunctions these
parameters are already set --- the values of $\beta$ were found individually 
for
each meson, and the quark masses were fitted: $m_u = 220$~MeV, $m_d = 220$~MeV,
and $m_s = 419$~MeV.  We have treated all mesons as narrow
resonances, and have ignored mass differences between members of the same
isospin multiplet \footnote{The one exception was for the decay 
$\phi \to K^+ K^-$ where the charged and neutral kaon mass
difference is significant to the phase space.}.  Masses were taken from the
Review of Particle Properties 1994 \cite{pdb} if the state was included in 
their Meson Summary Table \footnote{The one exception was the $1^3P_0$ 
$s \bar{s}$ state --- see Table \ref{table4}.}.  
If it was not, then the masses predicted in
Ref.~\cite{godfrey85} were used.  (This includes the masses of the $1^3F_2$
and $1^3F_4$ $s\bar{s}$ mesons: 2240~MeV and 2200~MeV
respectively.)  Meson flavour wavefunctions were also taken from
Ref.~\cite{godfrey85} - for the isoscalars we assumed ideal mixing ($\phi_{\rm
non strange} = \frac{1}{\sqrt{2}}(u \bar{u}+d \bar{d})$, $\phi_{\rm strange} =
s \bar{s}$), except for the radial ground state pseudoscalars, where we assumed
perfect mixing ($\phi_\eta = \frac{1}{\sqrt{2}}(\phi_{\rm non
strange}-\phi_{\rm strange})$, $\phi_{\eta'} = \frac{1}{\sqrt{2}}(\phi_{\rm non
strange}+\phi_{\rm strange})$).

We fitted $\gamma$, the one undetermined parameter of the model, in a global
least squares fit of 28 of the best known meson decays.  (We minimized the
quantity defined by $\chi^2 =\sum_i (\Gamma^{model}_i -
\Gamma^{exp}_i)^2/\sigma_{\Gamma_i}^2$ where $\sigma_{\Gamma_i}$ is the
experimental error \footnote{For the calculations in the flux-tube breaking
model, a 1\% error due to the numerical integration was added in quadrature
with the experimental error.}.)  The experimental values for these decays and
the fitted values for the six cases are listed in Table \ref{table2}.  To give
a more descriptive picture of the results we plotted in Fig.~\ref{figure2}, on
a logarithmic scale, the ratio of the fitted values to the experimental values.
From Table \ref{table2} one can see that the results for the $^3P_0$ and
flux-tube breaking models for the SHO wavefunctions are very
similar\footnote {The one exception to this is the S-wave decay 
$K^*_0(1430) \to K\pi$
which seems particularly sensitive to the model.}.  We therefore only plotted
the $^3P_0$ model results using the SHO wavefunctions and the flux-tube
breaking model results for the RQM wavefunctions.  A reference line is drawn in
each case for $\Gamma^{model}/\Gamma^{exp}=1$ to guide the eye.  Since all the
partial widths are proportional to $\gamma^2$, using a different fit strategy
rescales $\gamma$.  This is equivalent to simply shifting all points on the
plot simultaneously making it easy to visualize any change in agreement for
specific decays.

The KIPSN gives a better overall fit to the data.  Even so, certain decays,
$K_3^*(1780) \to K\rho$ and $f_4(2050) \to \omega\omega$ for example, are fit
much better using the RPSN.  For both the RPSN and KIPSN one can see in
Fig.~\ref{figure2} that a significant number of the decays differ from the
experimental values by factors of two or more.  Decays with two pseudoscalars
in the final state tend to do better with the KIPSN but the KIPSN generally
underestimates decays of high L mesons with vector mesons in the final states.
On the other hand the RPSN tends to overestimate decays with two pseudoscalars
in the final states.  Similar observations can be made for the flux-tube
breaking model using the RQM wavefunctions.  Having said all this we stress
that these are only general observations and exceptions can be found to any of
them in Table \ref{table2}.  One must therefore be very careful not to take the
predictions at face value but should try if possible to compare the predicted
decay to a similar one that is experimentally well known.

Finally, we consider the sensitivity of our results to $\beta$.  In addition to
the fits discussed above, we 
performed simultaneous fits of both $\gamma$ and $\beta$ to the 28 
decay widths for both the RPSN and the KIPSN.  The resulting values of 
$\gamma$ and $\beta$ are 13.4 and 481~MeV respectively for RPSN and 
5.60 and 371~MeV respectively for KIPSN.  In both cases the overall 
fits improved slightly, with some widths in better agreement and some 
in worse agreement with experiment when compared to the fits for 
$\beta=400$~MeV.  However, the fitted widths of the most relevant 
$^3F_4$ decays improve slightly for RPSN but show mixed results for 
KIPSN.  We also redid our fits of $\gamma$ to the
decay widths for $\beta=350$~GeV 
and $\beta=450$~GeV.   For $\beta=350$~MeV the overall fit improves 
slightly for KIPSN although the predicted $f_4(2050)$ decay widths
are a little worse and 
the $K_4(2045)$ widths are a little better.  For RPSN the overall fit is 
a little 
worse as are the $^3F_4$ decays.  For $\beta=450$~MeV the overall fit with 
KIPSN becomes a little worse as does the fitted $^3F_4$ widths while 
for RPSN the overall fit and fitted $^3F_4$ widths become a little
better.  
We conclude that while there is some sensitivity to $\beta$, the 
results for modest changes in $\beta$ (including the $\beta$ we obtain by
fitting $\gamma$ and $\beta$ simultaneously)
are consistent with those for $\beta=400$~MeV within the overall 
uncertainty we assign to our results.  It should be stressed that it 
is not sufficient to simply change $\beta$ but that a new value of 
$\gamma$ must be fitted to the experimental widths included in our 
fit.

\section{RESULTS FOR $^3F_2$ and $^3F_4$ $\lowercase{s}\bar{\lowercase{s}}$ 
MESON DECAYS}
\label{sec3}

Using the $\gamma$'s obtained from our fit we calculated all kinematically
allowed partial widths for the $^3F_2$ and $^3F_4$ $s\bar{s}$ meson decays.
The results are given in Tables \ref{table3} and \ref{table4}.

For the $^3F_4$ state the main decay modes are:
\begin{equation}
f_4' \to K^*(892) K^*(892), \; K\bar{K}, \; KK^*(892), \;
\phi\phi, \; K K_2^*(1430), \; K K_1(1400), \; \eta\eta, \; \eta \eta' 
\end{equation}

For the KIPSN and the SHO wavefunctions the total width is 132~MeV with the
$^3P_0$ model.  For this set of assumptions the $K\bar{K}$, $\eta\eta$, and
$\eta\eta'$ modes are probably reasonably good estimates.  However, the decay
widths to $K K^*(892)$ and $K^*(892) K^*(892)$ are likely to be larger than the
predictions.  On this basis it does not seem likely to us that the $f_4'$ width
is less than the predicted total width by a factor of two or more, i.e. we do
not expect it to be less than about 70~MeV.  If anything, we would expect it to
be larger than the predicted width, i.e. $ > 140$~MeV.

For the $^3F_2$ state we obtain results similar to the $^3F_4$ state for the
$K\bar{K}$, $KK^*(892)$, and $K^*(892) K^*(892)$ modes. However, the $^3F_2$
also has large partial widths to $K K_1(1270)$, $K^*(892) K_1 (1270)$, 
$K K_2^*(1430)$ and $\eta f_1(1510)$.  In fact, $K K_1(1270)$ is the 
dominant decay mode.  It
is large in all variations of the calculation we give in Table IV.  The most
closely related decay in our fit is the decay $\pi_2(1670) \to f_2(1270) \pi$
which is relatively large and is well reproduced by the KI normalization and
SHO wavefunction case.  The total width for this case is $\sim 400$~MeV
\footnote{We note that the LASS collaboration has observed a $K_2^*(1980)$
state with a large total width of $373\pm 33 \pm 60$~MeV which could be
associated with the strange meson partner of the $^3F_2 (s\bar{s})$ meson
\cite{pdb}}.  Even if this width
 is overestimated by a factor of two, it would
still be too large to identify with the $\xi(2220)$.

Although this result appears surprising it has a straightforward explanation.
Examining Table \ref{table4}, the lowest angular momentum final states in
$f_2'$ decay are P-waves.  All of these decays are relatively broad but the
$f_2' \to K_1(1270) K$ is the P-wave decay with the largest available
phase space.  In fact, one could almost order the P-wave decays using
phase space alone.  The analogous decay of the $f_4'$ is in an F-wave and
therefore is subject to a larger angular momentum barrier. The lowest angular
momentum partial wave for $f_4'$ decays is a D-wave which although it has the
largest partial width of all $f_4'$ decays is still smaller than the P-wave
$f_2'$ decay.

As another measure of the reliability of these predictions we calculated 
the widths of the $K_4^*(2045)$ and $f_4(2050)$ mesons (the $^3F_4$ $K$-like
and non-strange isovector mesons, respectively).  The results for all
significant kinematically allowed final states are given for the $^3P_0$ model 
using SHO wavefunctions in Tables \ref{table5} and \ref{table6} respectively.  
The results are consistent with the general fit results
given in Table \ref{table2} and Fig.~\ref{figure2}.
In general, the widths calculated using RPSN
tend to be larger and those calculated using the KIPSN 
tend to be smaller.  More specifically, decays to two pseudoscalar 
mesons using RPSN are generally overestimated while the results 
calculated using KIPSN are in reasonable agreement with 
experiment.  There is no pattern for the decays to two vector final 
states.  The decay $K_4^*(2045) \to K^*(892) \rho$ is greatly overestimated 
using RPSN but is in good agreement using KIPSN.  In 
contrast, the predicted decay $f_4(2050) \to \omega\omega$ agrees well 
using RPSN but is greatly underestimated using KIPSN.  
The total widths tend to be overestimated using RPSN but are 
underestimated using KIPSN, both to varying degrees.  The 
only conclusion we can draw from these results is that the total 
width probably lies between the two estimates but it is difficult to
guess if it is closer to the lower or upper value.

Finally, in Table VII we give the predicted total widths for the 
$^3F_2$ and $^3F_4$ $s\bar{s}$ states for the different values of 
$\beta$ considered in the previous section.  Although they vary 
considerably, by roughly a factor of 2 going from $\beta=350$~MeV to 
$\beta=450$~GeV (except for the $\Gamma(^3F_2)$ with RPSN which 
varies by a factor of 3), these values are consistent within the large 
uncertainties we assign to our results.


\section{DISCUSSION AND CONCLUSIONS}

The motivation for this paper was to re-examine the possibility that the
$\xi(2220)$ is an L=3 $s\bar{s}$ meson. This question is especially timely
given the recent BES measurements of a narrow resonance with a mass of 2.2~GeV
seen in $J/\psi$ radiative decays.  To do so we calculated all kinematically
allowed hadronic decays of the $^3F_2$ and $^3F_4$ $s\bar{s}$ states using
several variations of the flux-tube breaking decay model.

It appears very unlikely that the $\xi(2220)$ can be understood as the $^3F_2$
$s\bar{s}$ state.  All variations of our calculation indicate that the $^3F_2$
$s\bar{s}$ is rather broad, $\gtrsim 400$~MeV.  
The dominant decay mode is the
difficult to reconstruct $K K_1(1270)$ final state. Other final 
states with 
large branching ratios are 
$K^*(892) K_1(1270)$, $K K^*(892)$, $K^*(892) K^*(892)$, $K K_2^*(1430)$, 
$K\bar{K}$, and $\eta f_1(1510)$.

It is more likely that the $^3F_4(s\bar{s})$ state can be associated 
with the $\xi(2220)$.  The calculated width is $\sim 140$~MeV but 
given the uncertainties of the models it is possible, although 
perhaps unlikely,  that the 
width could be small enough to be compatible with the width 
reported in the Review of Particle Properties 1994 \cite{pdb}.  
In this scenario the 
largest decay modes are to $K^*(892) K^*(892)$, $K \bar{K}$, $K K^*(892)$, and 
$\phi\phi$.  Since only the $K\bar{K}$ final state has been observed 
an important test of this interpretation would be the observation of 
some of these other modes.

There are, however, some problems with the $^3F_4(s\bar{s})$ 
identification of the $\xi(2220)$.  Foremost is the flavour symmetric 
decay patterns recently measured by the BES collaboration 
\cite{bes}. These results contradict the expectations for a
conventional $s\bar{s}$ meson.  Second is the wide range of measured 
widths for this state.  Although the Review of Particle Properties 1994
lists an 
average width of $38^{+15}_{-13}$~MeV the widths measured in 
hadron production experiments, LASS and E147, are larger while those 
measured in $J/\psi$ radiative decay tend to be narrow.  The 
exception is the DM2 experiment which does not see, in $J/\psi$ 
radiative decay, a narrow state in $J/\psi \to \gamma K\bar{K}$ but 
does observe a relatively broad state at this mass.  

To account for these contradictions we propose a second explanation 
of what is being observed in this mass region --- that two different
hadron states are observed, a narrow state produced in 
$J/\psi$ radiative decay and a broader state produced in hadron beam 
experiments.  The broader state would be identified with the 
$^3F_4(s\bar{s})$ state.    The predicted width is consistent with 
the quark model predictions and the LASS collaboration shows 
evidence that its quantum numbers are $J^{PC}=4^{++}$.  We would 
then identify the narrow hadron state observed in the gluon rich 
$J/\psi$ radiative decays as a glueball candidate predicted by 
lattice gauge theory results \cite{lattice}. Recent lattice results 
indicate that glueballs may be narrower than one might naively 
expect \cite{lattice2}.  The scalar glueball width
is expected to be less that 200 MeV and one might expect a higher angular 
momentum state to be even narrower.
The narrow state is not seen in 
hadron beam production because it is narrow,
is produced weakly in these experiments through intermediate gluons, 
and is hidden by the 
$s\bar{s}$ state.  Conversely, the broader state is not seen in 
$J/\psi$ radiative decays since this mode preferentially produces 
states with a high glue content.  Crucial to this explanation is the 
experimental verification of the BES results on the flavour symmetric 
couplings of the state produced in $J/\psi$ radiative decay and the 
observation of other decay modes for the broader state in addition to the 
theoretical verification that the predicted tensor glueball is as 
narrow as the observed width.

The $\xi(2220)$ has been a longstanding source of controversy.  It is 
a dramatic reminder that there still is much that we don't 
understand about hadron spectroscopy and demonstrates the need for 
further experimental results to better understand this subject and 
ultimately better understand non-Abelian gauge theories, of which QCD 
is but one example.

\acknowledgments

This research was supported in part by the Natural Sciences and
Engineering Research Council of Canada. 
S.G. thanks Nathan Isgur and Eric Swanson for helpful conversations.


\appendix
\section{Review of the $^3P_0$ Model of Meson Decay}
\label{appa}

\begin{figure}
\centerline{\epsfig{file=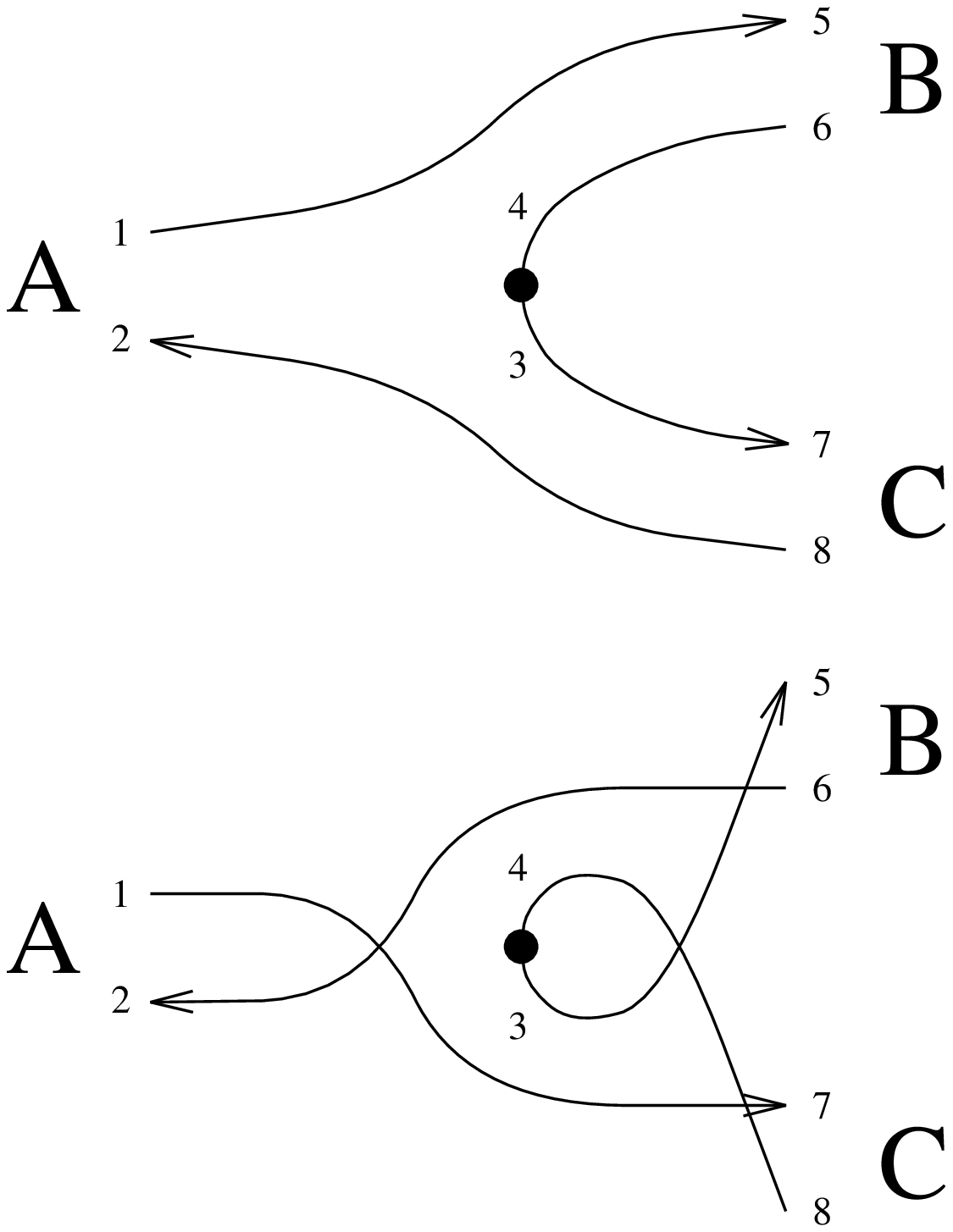,width=12.0cm,clip=}}
\caption{The two possible diagrams contributing to the meson decay $A
\rightarrow B C$ in the $^3P_0$ model.  In many cases only one of these
diagrams will contribute. \label{3p0decay}}
\end{figure}

\begin{figure}
\centerline{\epsfig{file=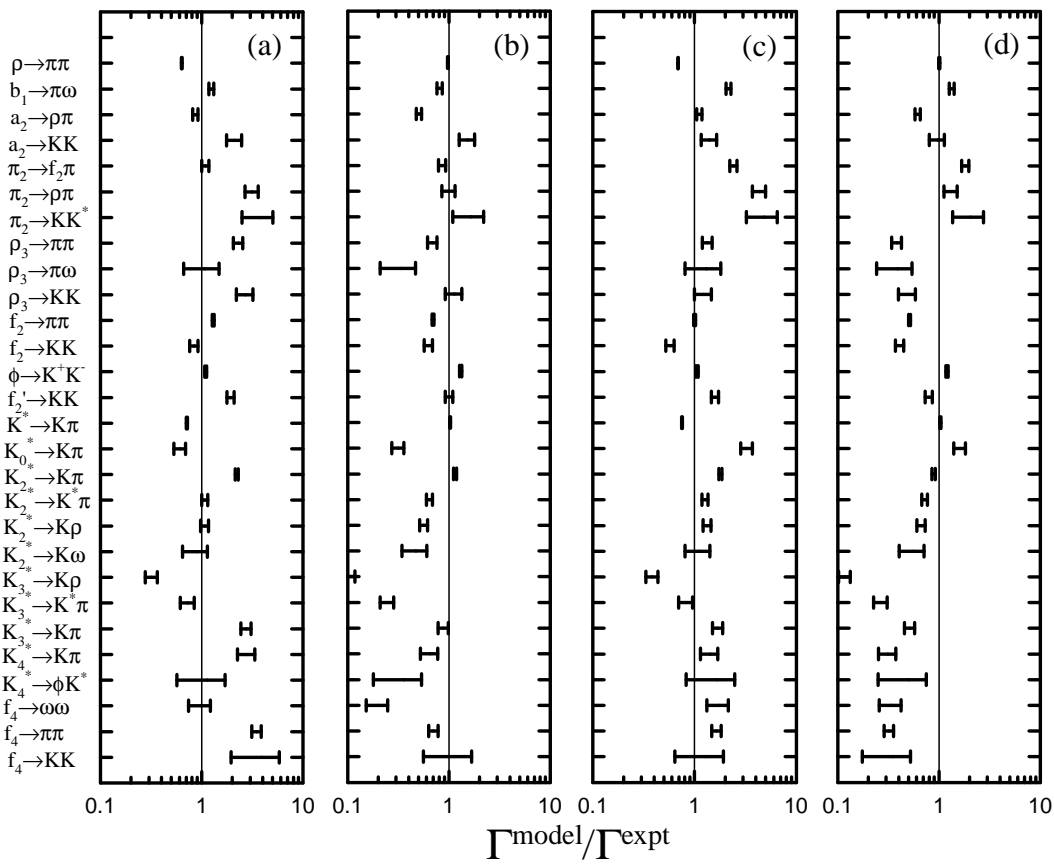,width=12.0cm,clip=}}
\caption[]{The ratio of decay model predictions for partial widths to 
the experimental values.  The error bars only include the effects of  
experimental errors in the ratios.  (a) and (b) correspond to the $^3P_0$ 
model using SHO wavefunctions and 
RPSN and KIPSN respectively
(columns 3 and 4 of Table \ref{table2}).  (c) and (d) correspond to the 
flux-tube breaking 
model using the RQM wavefunctions of Ref.~\cite{godfrey85} and
RPSN and KIPSN respectively
(columns 7 and 8 of Table \ref{table2}). \label{figure2}}
\end{figure}

We are looking at the meson decay $A \rightarrow B C$ in the $^3P_0$ model
(Fig.~\ref{3p0decay}).
Define the S matrix
\[
S = I -2\pi i \:\delta(E_f-E_i)\:T
\]
and then
\begin{equation}
\langle f|T|i\rangle = \delta^3(\vec{P}_f-\vec{P}_i)\: M^{M_{J_A} M_{J_B}
M_{J_C}} \label{texpectation}
\end{equation}
which gives, using relativistic phase space, the decay width in the centre of
mass (CM) frame
\begin{equation}
\Gamma = \pi^2 \frac{P}{M_A^2}\frac{s}{(2J_A+1)} \sum_{M_{J_A},M_{J_B},
M_{J_C}} |M^{M_{J_A} M_{J_B} M_{J_C}}|^2. \label{width}
\end{equation}
Here $P$ is the magnitude of the momentum of either outgoing meson, $M_A$ is
the mass of meson A, $|J_A,M_{J_A}\rangle$ are the quantum numbers of the total
angular momentum of A, $s \equiv 1/(1+\delta_{BC})$ is a statistical
factor which is needed if B and C are identical particles, and 
$M^{M_{J_A} M_{J_B} M_{J_C}}$ is the decay amplitude.

For the meson state we use a mock meson defined by\cite{mockmeson}:
\begin{eqnarray}
|A(n_A \mbox{}^{2S_A+1}L_A \,\mbox{}_{J_A M_{J_A}}) (\vec{P}_A)
\rangle &=& \sqrt{2 E_A}\: \sum_{M_{L_A},M_{S_A}}\! \langle L_A M_{L_A}
S_A M_{S_A} | J_A M_{J_A} \rangle \nonumber\\
&&\times \int\! {\rm d}^3\vec{p}_A\; \psi_{n_A L_A M_{L_A}}\!(\vec{p}_A)\;
\chi^{1 2}_{S_A M_{S_A}}\, \phi^{1 2}_A\;\, \omega^{1 2}_A \nonumber\\
&&\times |q_1({\scriptstyle \frac{m_1}{m_1+m_2}}\vec{P}_A+\vec{p}_A)\:
\bar{q}_2({\scriptstyle \frac{m_2}{m_1+m_2}}\vec{P}_A-\vec{p}_A)\rangle 
\label{mockmeson}
\end{eqnarray}
The subscripts 1 and 2 refer to the the quark and antiquark of meson A, 
respectively; $\vec{p}_1$ and $m_1$ are the momentum and mass of the quark.  
Note that the mock meson is normalized
relativistically to $2E_A\:\delta^3 (\vec{P}_A-\vec{P}_A')$, but uses
nonrelativistic spinors and CM coordinates ($\vec{P}_A = \vec{p}_1 +
\vec{p}_2$ is the momentum of the CM; $\vec{p}_A = (m_2 \vec{p}_1 -
m_1 \vec{p}_2)/(m_1 + m_2)$ is the relative momentum).
$n_A$ is the radial quantum number; $|L_A,M_{L_A}\rangle$ and 
$|S_A,M_{S_A}\rangle$ are the
quantum numbers of the orbital angular momentum between the two quarks, and 
their total spin angular momentum, respectively;  
$\langle L_A M_{L_A}
S_A M_{S_A} | J_A M_{J_A} \rangle$ is a Clebsch-Gordan coefficient.
$\chi^{1 2}_{S_A M_{S_A}}$, $\phi^{1 2}_A$ and $\omega^{1 2}_A$ are the
appropriate factors for combining the quark spins, flavours and colours,
respectively, and $\psi_{n_A L_A M_{L_A}}\!(\vec{p}_A)$ is the relative
wavefunction of the quarks in momentum space.  

For the transition operator we use
\begin{equation}
T = - 3 \gamma \sum_m\: \langle 1m\,1\!-\!\!m|00
\rangle\, \int\!{\rm d}^3\vec{p}_3\; {\rm d}^3\vec{p}_4
\delta^3(\vec{p}_3+\vec{p}_4)\:
{\cal Y}^m_1({\scriptstyle \frac{\vec{p}_3-\vec{p}_4}{2}})\;
\chi^{3 4}_{1 -\!m}\; \phi^{3 4}_0\;\, \omega^{3 4}_0\;
b^\dagger_3(\vec{p}_3)\; d^\dagger_4(\vec{p}_4) \label{tmatrix}
\end{equation}

where $\gamma$ is the one undetermined parameter in the model
\footnote{Our value of $\gamma$ is higher than that used by Kokoski and
Isgur \cite{kokoski87} by a factor of $\sqrt{96 \pi}$ due to different
field theory conventions, constant factors in 
$T$, etc.  The calculated values of the widths are, of course, unaffected.}  
and ${\cal Y}^m_l(\vec{p}) \equiv p^l
Y^m_l(\theta_p,\phi_p)$ is a solid harmonic that gives the
momentum-space distribution of the created pair.  Here the spins and
relative orbital angular momentum of the created quark and antiquark
(referred to by subscripts 3 and 4 respectively) are combined to give
the pair the overall $J^{PC}=0^{++}$ quantum numbers (in the $^3P_0$
state).

Combining Eq.~\ref{texpectation}, \ref{mockmeson} and \ref{tmatrix} gives for 
the amplitude in the CM frame (after doing the colour wavefunction 
overlap):
\begin{eqnarray}
M^{M_{J_A} M_{J_B} M_{J_C}}(\vec{P}) \;&=&\; \gamma\;\sqrt{8 E_A E_B E_C}
\sum_{\renewcommand{\arraystretch}{.5}\begin{array}[t]{l}
\scriptstyle M_{L_A},M_{S_A},M_{L_B},M_{S_B},\\
\scriptstyle M_{L_C},M_{S_C},m
\end{array}}\renewcommand{\arraystretch}{1}\!\!
\langle L_A M_{L_A} S_A M_{S_A} | J_A M_{J_A} \rangle \nonumber\\
&&\times \langle L_B M_{L_B} S_B M_{S_B} | J_B M_{J_B} \rangle
\langle L_C M_{L_C} S_C M_{S_C} | J_C M_{J_C} \rangle
\langle 1m\,1\!-\!\!m|00 \rangle \nonumber\\
&&\times \langle \chi^{1 4}_{S_B M_{S_B}} \chi^{3 2}_{S_C M_{S_C}} |
\chi^{1 2}_{S_A M_{S_A}} \chi^{3 4}_{1 -\!m} \rangle
\left[ \langle \phi^{1 4}_B \phi^{3 2}_C | \phi^{1 2}_A \phi^{3 4}_0 \rangle\:
I(\vec{P},m_1,m_2,m_3) \right.\nonumber\\
&&\left.+(-1)^{1+S_A+S_B+S_C}
\langle \phi^{3 2}_B \phi^{1 4}_C | \phi^{1 2}_A \phi^{3 4}_0 \rangle\:
I(-\vec{P},m_2,m_1,m_3) \right].  \label{amplitude}
\end{eqnarray}
The two terms
in the last factor correspond to the two possible diagrams in
Fig.~\ref{3p0decay} - in the first diagram the quark in A ends up B; in the 
second it ends up in C.
The momentum space integral $I(\vec{P},m_1,m_2,m_3)$ is given by
\begin{equation}
I(\vec{P},m_1,m_2,m_3) = \int\!{\rm d}^3\vec{p}\;
\psi^*_{n_B L_B M_{L_B}}\!
({\scriptstyle \frac{m_3}{m_1+m_3}}\vec{P}+\vec{p})\;
\psi^*_{n_C L_C M_{L_C}}\!
({\scriptstyle \frac{m_3}{m_2+m_3}}\vec{P}+\vec{p})\;
\psi_{n_A L_A M_{L_A}}\!
(\vec{P}+\vec{p})\;
{\cal Y}^m_1(\vec{p}) \label{integral}
\end{equation}
where we have taken $\vec{P} \equiv \vec{P_B} = - \vec{P_C}$.

\section{Review of the Flux-tube Breaking Model of Meson Decay}
\label{appb}

\begin{figure}
\centerline{\epsfig{file=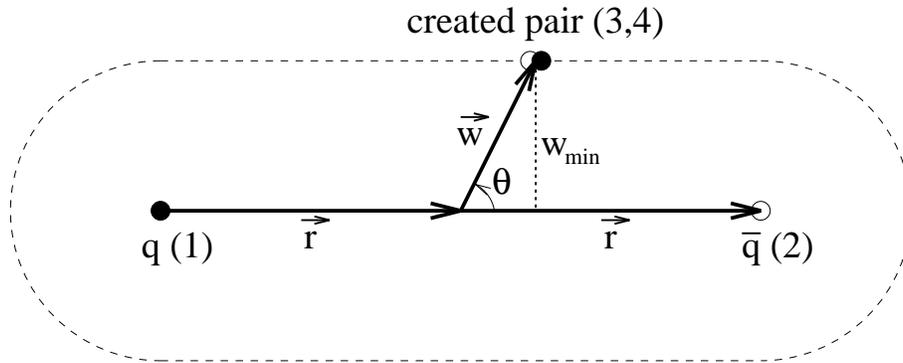,width=12.0cm,clip=}}
\caption{The position-space coordinates used in the flux-tube model.  The 
cigar-shaped dashed line shows a possible surface of constant 
$\omega_{\rm min}$.
\label{fluxvectors}}
\end{figure}

The flux-tube breaking model of meson decay extends the $^3P_0$ model 
by considering the
actual dynamics of the flux-tubes.  This is done by including a factor
representing the overlap of the flux-tube of the initial meson with those of
the two outgoing mesons.  Kokoski and Isgur \cite{kokoski87} have
calculated this factor by treating the flux-tubes as vibrating strings.  They
approximate the rather complicated result by replacing the undetermined
parameter $\gamma$ in the $^3P_0$ model with a function of the location of the
created quark-antiquark pair, and a new undetermined parameter $\gamma_0$:
\[
\gamma(\vec{r},\vec{w}) = \gamma_0 e^{-\frac{1}{2}b w_{\rm min}^2}.
\]
Here b is the string tension (a value of $0.18$~${\rm GeV}^2$ is typically
used)
and $w_{\rm min}$ is the shortest distance from the line segment connecting the
original quark and antiquark to the location at which the new quark-antiquark
pair is created from the vacuum (see Fig.~\ref{fluxvectors}):
\[
w_{\rm min}^2=\left\{ \begin{array}{ll}
w^2 \sin^2{\theta}, & \mbox{if $r \geq w\left|\cos{\theta}\right|$}\\
r^2+w^2-2rw \left|\cos{\theta}\right|, &
\mbox{if $r < w\left|\cos{\theta}\right|$}
\end{array}
\right.
\]

To incorporate this into the $^3P_0$ model, we first Fourier transform
Eq.~\ref{integral} so that the integral is over position-space.  We then pull
the parameter $\gamma$ inside the integral, and replace it by the function of
position $\gamma(\vec{r},\vec{w})$.  The expression for the amplitude in the
flux-tube model is then the same as that of Eq.~\ref{amplitude} except that
$\gamma$ is replaced by $\gamma_0$, and $I(\vec{P},m_1,m_2,m_3)$ is replaced by
\begin{eqnarray*}
I^{ft}(\vec{P},m_1,m_2,m_3)&=&-\frac{8}{(2\pi)^{\frac{3}{2}}}
\int\!{\rm d}^3\vec{r}\int\!{\rm d}^3\vec{w}\;\:
\psi^*_{n_B L_B M_{L_B}}\!
(-\vec{w}-\vec{r})\;
\psi^*_{n_C L_C M_{L_C}}\!
(\vec{w}-\vec{r})\\
&&\times{\cal Y}^m_1\left(\left[\left(\vec{P}+i\vec{\nabla}_{\vec{r}_A}\right)
\psi_{n_A L_A M_{L_A}}\!(\vec{r}_A)\right]_{\vec{r}_A=-2\vec{r}}\right)
e^{-\frac{1}{2}b w_{\rm min}^2} \nonumber\\
&&\times e^{i\vec{P}\cdot\left[\vec{r} \left(\frac{m_1}{m_1+m_3}+
\frac{m_2}{m_2+m_3}\right) + \vec{w}\left(\frac{m_1}{m_1+m_3}-
\frac{m_2}{m_2+m_3}\right)\right]}
\end{eqnarray*}
where the $\psi$'s are now the relative wavefunctions in position space.

\section{Converting to Partial Wave Amplitudes}
\label{appc}

The decay amplitudes of the $^3P_0$ and flux-tube breaking models derived in
Appendices \ref{appa} and \ref{appb}, $M^{M_{J_A} M_{J_B} M_{J_C}}$, are given
for a particular basis of the final state: $|\theta,\phi,M_{J_B},M_{J_C}\rangle
\equiv |\Omega,M_{J_B},M_{J_C}\rangle$.  Here $\theta$ and $\phi$ are the
spherical polar angles of the outgoing momentum of meson B in the CM frame.

We would prefer to calculate amplitudes for particular partial waves, since
they are what are measured experimentally: $|J,M,S,L\rangle$.  Here
$|J,M\rangle$ are the quantum numbers of the total angular momentum of the
final state, $|S,M_S\rangle$ are the quantum numbers for the sum of the
total angular momenta of B and C, and $|L,M_L\rangle$ are the quantum numbers 
for the orbital angular momentum between B and C.

The formula for the decay width in terms of partial wave amplitudes is 
different from Eq.~\ref{width}:
\[
\Gamma = \sum_{S,L} \Gamma^{S L}
\]
where
\[
\Gamma^{S L} = \frac{\pi}{4} \frac{P s}{M_A^2} |M^{S L}|^2.
\]
$M^{S L}$ is a partial wave amplitude, and $\Gamma^{S L}$ is the partial width 
of that partial wave.

We used two methods to convert our calculated amplitudes to the partial wave
basis \cite{convert}: a recoupling calculation, and by use of the
Jacob-Wick Formula.

\subsection{Converting by a Recoupling Calculation}

The result of a recoupling calculation is
\begin{equation}
M^{S L}(P) = \sum_{M_{J_B},M_{J_C},M_S,M_L} \!\!\!
\langle L M_L S M_S|J_A M_{J_A} \rangle 
\langle J_B M_{J_B} J_C M_{J_C} | S M_S \rangle 
\int\!{\rm d}\Omega\; 
Y^*_{L M_L}\!(\Omega)\; M^{M_{J_A} M_{J_B} M_{J_C}}(\vec{P}). 
\label{recoupling}
\end{equation}
Note that this can be done for any value of $M_{J_A}$; alternatively, one could
sum over $M_{J_A}$ and divide by $(2 J_A +1)$, on the right side.

\subsection{Converting with the Jacob-Wick Formula}

The Jacob-Wick formula relates the partial wave basis $|J,M,S,L\rangle$ to the 
helicity basis $|J,M,\lambda_B,\lambda_C\rangle$, where $\lambda_B$ and 
$\lambda_C$ are the helicities of B and C, respectively.  To use it we must 
first convert the basis that we calculate with to the helicity basis.  
This is done by first choosing $\vec{P} \equiv \vec{P_B}$ to lie 
along the positive z
axis (in the CM frame still), so that $\lambda_B = M_{J_B}$ and $\lambda_C = -
M_{J_C}$.  Then one can use another expression that relates the helicity basis
to the basis $|\Omega,\lambda_B,\lambda_C\rangle$.

The final result is
\begin{eqnarray*}
M^{S L}(P) &=& \frac{\sqrt{4 \pi (2 L+1)}}{2 J_A +1} \!\!
\sum_{M_{J_B},M_{J_C}} 
\langle L 0 S (M_{J_B}+M_{J_C})|J_A (M_{J_B}+M_{J_C})\rangle 
\langle J_B M_{J_B} J_C M_{J_C} | S (M_{J_B}+M_{J_C}) \rangle \\
&&\times M^{(M_{J_A}=M_{J_B}+M_{J_C}) M_{J_B} M_{J_C}}(P \hat{z}).
\end{eqnarray*}
Here $M_{J_A}$ in the calculated amplitude is replaced by $M_{J_B}+M_{J_C}$.

\section{Calculational Techniques}
\label{appd}

The decay amplitudes in the $^3P_0$ model were converted to partial wave
amplitudes by means of a recoupling calculation.  The whole expression for the
amplitudes, including the integrals of Eqs.~\ref{integral} and
\ref{recoupling}, was converted into a sum over angular momentum quantum
numbers, using the techniques of Roberts and Silvestre-Brac \cite{roberts} (a
result very similar to theirs was obtained).  These techniques require that the
radial portion of the meson wavefunctions be expressible in certain functional
forms, which encompass simple harmonic oscillator wavefunctions.  Our simple
wavefunctions obviously meet these requirements, and since the detailed 
wavefunctions of Ref.~\cite{godfrey85} are expansions in terms of SHO 
wavefunctions, they do too.

These expressions for the amplitudes were then computed symbolically using
routines written for Mathematica \cite{wolfram}.  These routines are usable
for any meson decay where the radial portion of the wavefunctions can be
expanded in terms of SHO wavefunctions, and are limited only by the size of the
symbolic problem that results, and the available computer resources.

In the flux-tube breaking model there are two 3-dimensional 
integrations
before converting to partial wave amplitudes.  The wish to be able to
write general routines for any meson decay meant that only two of the six
integrals could be done analytically; the remaining four must be done
numerically.  In order to minimize the numerical integration, the Jacob-Wick
formula, rather than a recoupling calculation, was used to convert to 
partial wave amplitudes since no further integrals are involved.

An integrand for each partial wave amplitude was prepared symbolically and
converted to Fortran code using routines written for Mathematica, and then
integrated numerically using either adaptive Monte Carlo (VEGAS \cite{vegas})
or a combination of adaptive Gaussian quadrature routines.  Again, these
routines are usable for any meson decay where the radial portion of the
wavefunctions can be expanded in terms of SHO wavefunctions, and are limited
only by the size of the problem and available computer resources.


\newpage

\newpage

\newpage
\begin{table}
\caption{Summary of $\xi(2220)$ measurements.}
\label{table1}
\begin{tabular}{llllll}
Experiment & Mass & Width & Production & Decays  \\
	& (MeV)	& (MeV) & 	& 	\\
\tableline
Mark III \tablenotemark[1] & $2231 \pm 8$ & $21 \pm 17$ & 
	$J/\psi \to \gamma \xi $ &
	$BR(J/\psi \to \gamma \xi) \times BR(\xi \to K^+ K^-)$ \\
	& & & & \qquad $= (4.2 ^{+1.7}_{-1.4}\pm 0.8)\times 10^{-5}$ \\
	&	&	&	&
	$BR(J/\psi \to \gamma \xi) \times BR(\xi \to K_S K_S)$ \\
	& & & & \qquad = $(3.1 ^{+1.6}_{-1.3}\pm 0.7)\times 10^{-5}$ \\
	&	&	&	&
	$BR(J/\psi \to \gamma \xi) \times BR(\xi \to \pi \pi)$ \\
	& & & & \qquad $< 2 \times 10^{-5}$ (90\% C.L.) 	 \\
	&	&	&	&
	$BR(J/\psi \to \gamma \xi) \times BR(\xi \to p\bar{p})$ \\ 
	& & & & \qquad $ < 2 \times 10^{-5}$ (90\% C.L.)	 \\
DM2 \tablenotemark[2] & $2230$ \tablenotemark[3] & $26$ \tablenotemark[3] & 
	$J/\psi \to \gamma \xi $ &
	$BR(J/\psi \to \gamma \xi) \times BR(\xi \to K^+ K^-) $ \\
	& & & & \qquad $ < 2.3 \times 10^{-5}$ 	(95\% C.L.) \\
	&	&	&	&
	$BR(J/\psi \to \gamma \xi) \times BR(\xi \to K_S K_S) $ \\
	& & & & \qquad $< 1.6 \times 10^{-5}$ (95\% C.L.)	 \\
	& $2197 \pm 17$ & $201 \pm 51$ & $J/\psi \to \gamma X  $ &
	$BR(J/\psi \to \gamma X) \times BR(X \to K_S K_S) $ \\
	& & & & \qquad $ \simeq  1.5 \times 10^{-4}$  \\
BES \tablenotemark[4] & $2233\pm 5$ & $19 \pm 11 $ & $J/\psi \to \gamma \xi $ &
	$BR(J/\psi \to \gamma \xi) \times BR(\xi \to \pi^+ \pi^-) $ \\
	& & & & \qquad $ = (5.6 ^{+1.8}_{-1.6}\pm 1.4)\times 10^{-5}$ 	 \\
	&	&	&	&
	$BR(J/\psi \to \gamma \xi) \times BR(\xi \to p\bar{p}) $\\
	& & & & \qquad $= (1.5 ^{+0.6}_{-0.5}\pm 0.5)\times 10^{-5}$ 	 \\
	&	&	&	&
	$BR(J/\psi \to \gamma \xi) \times BR(\xi \to K^+ K^-) $ \\
	& & & & \qquad $ = (3.3 ^{+1.6}_{-1.3}\pm 1.1)\times 10^{-5}$ 	 \\
	&	&	&	&
	$BR(J/\psi \to \gamma \xi) \times BR(\xi \to K_S K_S) $ \\
	& & & & \qquad $ = (2.7 ^{+1.1}_{-0.9}\pm 1.0)\times 10^{-5}$ 	 \\
LASS \tablenotemark[5] & $2209^{+17}_{-15}\pm 10$ & $60^{+107}_{-57}$ 
	& $K^- p \to K^- K^+ \Lambda$ &  \\
E147 \tablenotemark[6] & $2230 \pm 20$ & $80 \pm 30 $ 
	& $\pi^- p \to K_S K_S n$ &  	\\
PS185 \tablenotemark[7] & $2231$ \tablenotemark[3] & $30$ \tablenotemark[3] 
	& $p\bar{p} \to K_S K_S$ &
	$BR(\xi \to p\bar{p} ) \times BR(\xi \to K_S K_S) $ \\
	& & & & \qquad $< 5.4 \times 10^{-4}$ 	(3 S.D. J=4) \\
\end{tabular}
\tablenotetext[1]{Ref. \cite{markiii}}
\tablenotetext[2]{Ref. \cite{dm2}}
\tablenotetext[3]{Note that these values aren't measurements - they were
assumed in order to set the BR limits.}
\tablenotetext[4]{Ref. \cite{bes}}
\tablenotetext[5]{Ref. \cite{lass}}
\tablenotetext[6]{Ref. \cite{itep}}
\tablenotetext[7]{Ref. \cite{ps185}}
\end{table}

\newpage
\begin{table}
\caption{Experimental and calculated widths (in MeV) of decays used in our 
global fit of the decay models' parameters.}
\label{table2}
\begin{tabular}{llcccccc}
Decay & $\Gamma$(experiment) 
	& \multicolumn{2}{c}{$^3P_0$}
	& \multicolumn{2}{c}{Flux-Tube Breaking} 
	& \multicolumn{2}{c}{Flux-Tube Breaking} \\
	&	& \multicolumn{2}{c}{(SHO)}
	& \multicolumn{2}{c}{(SHO)}
	& \multicolumn{2}{c}{(RQM)} \\ 
	&	& RPSN & KIPSN & RPSN & KIPSN & RPSN & KIPSN \\
\tableline
$\gamma$ &  & 9.73 & 6.25 & 16.0 & 10.4 & 20.5 & 12.8 \\
\tableline
$\rho\to \pi \pi$ & $151.2 \pm 1.2$ & 96 & 148 & 93 & 148 & 104 & 152 \\
$b_1(1235)\to \omega \pi$ & $142 \pm 8$  & 176 & 115 & 155 & 104 & 
		306 & 190 \\
$a_2(1320) \to \rho\pi$ & $75.0 \pm 4.5$ & 65  & 38 & 67 & 40 & 84 & 46 \\
$a_2(1320) \to K\bar{K}$ & $5.2 \pm 0.9$  & 11 & 8.0 & 11 & 8.5 
		& 7.3 & 5.0 \\
$\pi_2(1670) \to f_2(1270) \pi$ & $ 135 \pm 11$ & 147 & 116 & 143 & 
		117 & 327 & 246 \\
$\pi_2(1670) \to \rho \pi$ & $ 74 \pm 11$ & 232 & 74 & 226 & 74 & 
		323 & 97 \\
$\pi_2(1670) \to K^*(892) \bar{K} +c.c.$ & $10.1 \pm 3.4$ & 38 & 17 & 37 & 
		17 & 49 & 21 \\
$\rho_3(1690) \to \pi \pi $ & $50.7 \pm 5.5$ & 116 & 35 & 122 & 38 & 
		68 & 19 \\
$\rho_3(1690) \to \omega \pi $ & $34 \pm 13$ & 36 & 11 & 39 & 13 & 
		45 & 13 \\
$\rho_3(1690) \to K \bar{K} $ & $3.4 \pm 0.6$  & 9.2 & 3.8 & 9.7 & 
		4.2 & 4.2 & 1.7 \\
$f_2(1270) \to \pi \pi$ & $156.8 \pm 3.2$ & 203 & 109  & 209 & 116 & 157 & 
	80 \\
$f_2(1270) \to K\bar{K}$ & $8.6 \pm 0.8 $ & 7.2 & 5.4 & 7.4 & 5.7 & 5.0 & 
	3.5 \\
$\phi \to K^+ K^-$ & $2.17 \pm 0.05$ & 2.37 & 2.83 & 2.28 & 2.80 & 
	2.30 & 2.60\\
$f_2'(1525) \to K\bar{K}$ & $61 \pm 5$  & 117 & 61 & 118 & 64 & 98 & 49 \\
$K^*(892) \to K\pi$ & $50.2 \pm 0.5 $ &  36 & 52 & 34 & 51 & 38 & 52 \\
$K_0^*(1430)\to K\pi$ & $267 \pm 36$ & 163 & 84 & 117 & 63 & 875 & 430 \\
$K_2^*(1430)\to K\pi$ & $48.9 \pm 1.7 $  & 108 & 56 & 112 & 60 & 88 & 43 \\
$K_2^*(1430)\to K^*(892)\pi$ & $24.8\pm 1.7$ & 27 & 16 & 27 & 17 & 31 & 18\\
$K_2^*(1430) \to K\rho$ & $8.7\pm 0.8$ & 9.3 & 4.9 & 9.6 & 5.2 & 12 & 5.8 \\
$K_2^*(1430) \to K\omega$ & $2.9 \pm 0.8$ & 2.6 & 1.4 & 2.6 & 1.4 & 
	3.2 & 1.6 \\
$K_3^*(1780)\to K \rho$ & $74 \pm 10$ & 24 & 7.7 & 25 & 8.4 & 28 & 8.7 \\
$K_3^*(1780)\to K^*(892)\pi$ & $45 \pm 7$ & 33 & 11 & 34 & 12 & 37 & 12 \\
$K_3^*(1780)\to K\pi$ & $31.7\pm 3.7$ & 87 & 28 & 92 & 30 & 54 & 16 \\
$K_4^*(2045)\to K\pi$ & $19.6\pm 3.8$ & 55 & 13 & 59 & 14 & 28 & 6.2 \\
$K_4^*(2045)\to K^*(892) \phi$ & $2.8 \pm 1.4$ & 3.2 & 1.0 & 3.3 & 
		1.1 & 4.7 & 1.4 \\
$f_4 (2050) \to \omega\omega$ & $54 \pm 13 $ & 53 & 11 & 54 & 11 & 
		94  & 18 \\
$f_4 (2050) \to \pi\pi$ & $35.4\pm 3.8$ & 123 & 25 & 132 & 28 & 58 & 11 \\
$f_4 (2050) \to K\bar{K}$ & $1.4 \pm 0.7$ & 5.4 & 1.6 & 5.8 & 1.7 & 
		1.8 & 0.5 \\
\end{tabular}
\end{table}

\newpage
\begin{table}
\caption{Calculated partial decay widths (in MeV) for the $^3F_4$ $s\bar{s}$ 
state.  We have calculated the
widths of all kinematically allowed decays, but only show those partial widths 
that are $\geq 1$~MeV in at least one model.  For this reason the 
total widths may
not equal the sum of the partial widths shown.  The subscripts on the 
decays refer to
the $S$ and $L$ (see Appendix \protect\ref{appc}) of the given partial wave 
--- the $L$ is in spectroscopic notation ($S, P, D, F, G, H$).}
\label{table3}
\begin{tabular}{lcccccc}
	Decay  & \multicolumn{2}{c}{$^3P_0$}
		& \multicolumn{2}{c}{Flux-Tube Breaking}
		& \multicolumn{2}{c}{Flux-Tube Breaking}\\
	& \multicolumn{2}{c}{(SHO)}
	& \multicolumn{2}{c}{(SHO)}
	& \multicolumn{2}{c}{(RQM)} \\ 
		& RPSN & KIPSN & RPSN & KIPSN & RPSN & KIPSN \\
\tableline
$f'_4 \to [K \bar{K}]_{0,G} $ & 118 & 29 & 125 & 31 & 62 & 14 \\
$f'_4 \to [K_r \bar{K}+c.c.]_{0,G} \tablenotemark[1]$ & 0.7 & 0.4 & 0.4 & 0.2 &
	2.4 & 1.2 \\
$f'_4 \to [K^*(892) \bar{K} +c.c.]_{1,G} $ & 107 & 27 & 115 & 29 & 112 & 26 \\
$f'_4 \to [K^* (1410) \bar{K} +c.c.]_{1,G} \tablenotemark[2]$ & 1.7 & 0.9 & 
	0.8 & 0.4 & 5.0 & 2.4\\
$f'_4 \to [K_1(1270) \bar{K}+c.c.]_{1,F} \tablenotemark[3]$ & 6.4 & 2.8 & 7.0 &
 	3.1 & 10 & 4.2 \\
$f'_4 \to [K_1(1270) \bar{K}+c.c.]_{1,H} \tablenotemark[3]$ & 1.3 & 0.6 & 
	1.4 & 0.6 & 3.7 & 1.5 \\
$f'_4 \to [K_1(1400) \bar{K}+c.c.]_{1,F} \tablenotemark[3]$ & 14 & 6.4 & 15 & 
	7.0 & 29 & 12 \\
$f'_4 \to [K_2^*(1430) \bar{K}+c.c.]_{2,F} $ & 15 & 7.0 & 16 & 7.7 & 35 & 15 \\
$f'_4 \to [K^*(892) \bar{K}^*(892)]_{0,G} $ & 2.1 & 0.5 & 2.3 & 0.6 & 4.3 & 
	1.0 \\
$f'_4 \to [K^*(892) \bar{K}^*(892)]_{2,D} $ & 181 & 44 & 184 & 46 & 312 & 72 
\\
$f'_4 \to [K^*(892) \bar{K}^*(892)]_{2,G} $ & 8.2 & 2.0 & 8.9 & 2.2 & 17 & 
3.9 \\
$f'_4 \to [\eta \eta]_{0,G} $ & 14 & 3.5 & 15 & 3.9 & 5.0 & 1.2 \\
$f'_4 \to [\eta' \eta]_{0,G} $ & 6.9 & 1.7 & 7.5 & 1.9 & 2.4 & 0.6 \\
$f'_4 \to [\phi \phi ]_{2,D} $ & 20 & 6.6 & 21 & 7.1 & 31 & 9.5 \\
\tableline
$\sum_i \Gamma_i$ & 498 & 132 & 522 & 142 & 633 & 166
\end{tabular}
\tablenotetext[1]{$K_r$ is our notation for the first radial excitation 
($2^1S_0$) of the $K$.}
\tablenotetext[2]{We used the following mixing \cite{godfrey85}:  
$\left\{
\begin{array}{lll} 
K^*(1410) &=& 1.00 (2^3S_1) + 0.04 (1^3D_1) \\
K^*(1680) &=& -0.04 (2^3S_1) + 1.00(1^3D_1)
\end{array}
\right.$}
\tablenotetext[3]{We used the following mixing 
\cite{pdb}:  
$\left\{
\begin{array}{lll} 
K_1(1270)^+ &=& \cos{45^\circ} (1^1P_1)^+ + \sin{45^\circ} (1^3P_1)^+ \\
K_1(1400)^+ &=& -\sin{45^\circ} (1^1P_1)^+ + \cos{45^\circ} (1^3P_1)^+
\end{array}
\right.$}
\end{table}

\newpage
\begin{table}
\caption{Calculated partial decay widths (in MeV) for the $^3F_2$ $s\bar{s}$
state.  We do not include a decay to $f_0(980) f_0(980)$ because we question 
its assignment as a $^3P_0$ $q \bar{q}$ meson.  At a more likely mass for the 
$^3P_0$ $s \bar{s}$ meson, this 
decay is kinematically inaccessible.  For other comments and notes, see 
Table~\protect\ref{table3}.}
\label{table4}
\begin{tabular}{lcccccc}
	Decay  & \multicolumn{2}{c}{$^3P_0$}
		& \multicolumn{2}{c}{Flux-Tube Breaking}
		& \multicolumn{2}{c}{Flux-Tube Breaking}\\
	& \multicolumn{2}{c}{(SHO)}
	& \multicolumn{2}{c}{(SHO)}
	& \multicolumn{2}{c}{(RQM)} \\ 
		& RPSN & KIPSN & RPSN & KIPSN & RPSN & KIPSN \\
\tableline
$f'_2 \to [K \bar{K}]_{0,D} $ & 51 & 12 & 47 & 12 & 101 & 23 \\
$f'_2 \to [K_r \bar{K}+c.c.]_{0,D} $ & 2.9 & 1.5 & 0.9 & 0.5 & 25 & 12 \\
$f'_2 \to [K^*(892) \bar{K} +c.c.]_{1,D} $ & 108 & 26 & 107 & 26 & 165 & 38 \\
$f'_2 \to [K^*(1410) \bar{K}+c.c.]_{1,D} $ & 2.6 & 1.3 & 0.6 & 0.3 
		& 4.0 & 1.9 \\
$f'_2 \to [K_1(1270) \bar{K}+c.c.]_{1,P} $ & 445 & 187 & 449 & 194 
		& 1072 & 426 \\
$f'_2 \to [K_1(1270) \bar{K}+c.c.]_{1,F} $ & 25 & 11 & 27 & 12 & 
		41 & 16 \\
$f'_2 \to [K_1(1400) \bar{K}+c.c.]_{1,P} $ & 14 & 6.3 & 15 & 6.9 & 29 & 12 \\
$f'_2 \to [K_1(1400) \bar{K}+c.c.]_{1,F} $ & 0.8 & 0.4 & 1.0 & 0.4 & 
		$\sim 0$ & $\sim 0$ \\
$f'_2 \to [K_2^*(1430) \bar{K}+c.c.]_{2,P} $ & 54 & 24 & 55 & 25 & 
		112 & 47 \\
$f'_2 \to [K_2^*(1430) \bar{K}+c.c.]_{2,F} $ & 9.6 & 4.3 & 10 & 4.7 & 
		22 & 9.1 \\
$f'_2 \to [K^*(892) \bar{K}^*(892)]_{0,D} $ & 24 & 5.7 & 24 & 5.9 & 39 & 8.9 \\
$f'_2 \to [K^*(892) \bar{K}^*(892)]_{2,D} $ & 14 & 3.3 & 14 & 3.4 & 23 & 5.1 \\
$f'_2 \to [K^*(892) \bar{K}^*(892)]_{2,G} $ & 48 & 12 & 52 & 13 & 83 & 19 \\
$f'_2 \to [K_1(1270) \bar{K}^*(892) +c.c.]_{1,P} $ & 99 & 40 & 102 & 
		42 & 209 & 79 \\
$f'_2 \to [K_1(1270) \bar{K}^*(892) +c.c.]_{1,F} $ & 0.5 & 0.2 & 0.6 & 0.2 & 
1.1 & 0.4 \\

$f'_2 \to [K_1(1270) \bar{K}^*(892) +c.c.]_{2,P} $ & 33 & 13 & 34 & 14 
		& 70 & 26 \\
$f'_2 \to [K_1(1270) \bar{K}^*(892) +c.c.]_{2,F} $ & 0.8 & 0.3 & 0.9 & 0.4 & 
	1.8 & 0.7 \\
$f'_2 \to [\eta \eta]_{0,D} $ & 14 & 3.3 & 13 & 3.2 & 20 & 4.4 \\
$f'_2 \to [\eta' \eta]_{0,D} $ & 29 & 7.0 & 29 & 7.2 & 29 & 6.6 \\
$f'_2 \to [f_1(1510) \eta]_{1,P} $ & 45 & 22 & 46 & 24 & 92 & 43 \\
$f'_2 \to [f_2'(1525) \eta]_{2,P} $ & 14 & 6.9 & 14 & 7.3 & 29 & 14  \\
$f'_2 \to [\eta' \eta']_{0,D} $ & 6.6 & 1.6 & 6.7 & 1.7 & 4.9 & 1.1 \\
$f'_2 \to [\phi \phi ]_{0,D} $ & 3.9 & 1.2 & 3.9 & 1.3 & 5.5 & 1.6 \\
$f'_2 \to [\phi \phi ]_{2,D} $ & 2.2 & 0.7 & 2.3 & 0.7 & 3.1 & 0.9 \\
$f'_2 \to [\phi \phi ]_{2,G} $ & 1.0 & 0.3 & 1.0 & 0.3 & 1.1 & 0.3 \\
\tableline
$\sum_i \Gamma_i$ & 1046 & 391 & 1058 & 406 & 2181 & 797 
\end{tabular}
\end{table}

\newpage
\begin{table}
\caption{Calculated partial decay widths (in MeV) for the $K_4^*(2045)$ state. 
For comments and additional notes, see Table~\protect\ref{table3}.}
\label{table5}
\begin{tabular}{lccc}
	Decay & Experiment & \multicolumn{2}{c}{$^3P_0$} \\
	&	& \multicolumn{2}{c}{(SHO)} \\
	&	&  RPSN & KIPSN \\
\tableline
$K_4^*(2045) \to [K \pi]_{0,G}$ & $19.6 \pm 3.8 $& 55 & 13 \\
$K_4^*(2045) \to [K \rho]_{1,G}$ & & 19 &  4.4 \\
$K_4^*(2045) \to [K b_1(1235)]_{1,F}$ & & 4.9 & 2.2 \\
$K_4^*(2045) \to [K a_1(1260)]_{1,F}$ & & 1.3 &  0.6 \\
$K_4^*(2045) \to [K a_2(1320)]_{2,F}$ & & 2.2 &  1.0 \\
$K_4^*(2045) \to [K^*(892) \pi]_{1,G}$ & & 23 & 5.5 \\
$\!\!\!\left. \begin{array}{l}
K_4^*(2045) \to [K^*(892) \rho]_{2,D} \\
K_4^*(2045) \to [K^*(892) \rho]_{2,G}
\end{array} \right\} $ &
$18 \pm 10$ \tablenote{This number is actually for the final state $K^*(892) 
\pi\pi$, 
and is the total for all partial waves.}&
$\begin{array}{c}
76 \\
2.1
\end{array} $ &
$\begin{array}{c}
18 \\
0.5
\end{array} $ \\
$K_4^*(2045) \to [K_1(1270) \pi]_{1,F}$ & & 1.6 & 0.7 \\
$K_4^*(2045) \to [K_1(1400) \pi]_{1,F}$ & & 5.3 & 2.6 \\
$K_4^*(2045) \to [K_2^*(1430) \pi]_{2,F}$ & & 5.2  & 2.6 \\
$K_4^*(2045) \to [K\eta']_{0,G}$ & & 3.3  & 0.9 \\
$K_4^*(2045) \to [K\omega]_{1,G}$ & & 6.0  & 1.4 \\
$K_4^*(2045) \to [K\phi]_{1,G}$ & & 1.1  & 0.4 \\
$K_4^*(2045) \to [K h_1(1170)]_{1,F}$ & & 2.9 & 1.3 \\
$K_4^*(2045) \to [K f_2(1270)]_{2,F}$ & & 1.3 & 0.6 \\
$K_4^*(2045) \to [K^*\eta]_{1,G}$ & & 4.9 & 1.4 \\
$K_4^*(2045) \to [K^*(892) \omega]_{2,D}$ & & 24 & 5.7 \\
$K_4^*(2045) \to [K^*(892) \phi]_{2,D}$ & $2.8 \pm 1.4$ & 3.2 & 1.0 \\
\tableline
$\sum_i\Gamma_i$ & $198 \pm 30$ & 247 & 65 
\end{tabular}
\end{table}

\newpage
\begin{table}
\caption{Calculated partial decay widths (in MeV) for the $f_4(2050)$ state. 
For comments and additional notes, see Table~\protect\ref{table3}.}
\label{table6}
\begin{tabular}{lccc}
	Decay & Experiment & \multicolumn{2}{c}{$^3P_0$} \\
	&	& \multicolumn{2}{c}{(SHO)} \\
	&	&  RPSN & KIPSN \\
\tableline
$f_4(2050) \to [\pi \pi]_{0,G}$ & $35.4\pm 3.8$ & 123 & 25 \\
$f_4(2050) \to [\pi \pi(1300)]_{0,G}$ & & 3.9 & 1.9 \\
$f_4(2050) \to [\pi a_1(1260)]_{1,F}$ & & 18 &  7.5 \\
$f_4(2050) \to [\pi a_2(1320)]_{2,F}$ & & 44 & 19 \\
$f_4(2050) \to [\pi \pi_2(1670)]_{2,D}$ &  & 2.1 & 1.8 \\
$f_4(2050) \to [\rho \rho]_{0,G}$ & & 1.9 & 0.4 \\
$f_4(2050) \to [\rho \rho]_{2,D}$ & & 159 & 33 \\
$f_4(2050) \to [\rho \rho]_{2,G}$ & & 7.3 & 1.5 \\
$f_4(2050) \to [\eta\eta]_{0,G}$ & & 3.2 & 0.9 \\
$f_4(2050) \to [\eta\eta']_{0,G}$ & & 1.0 & 0.3 \\
$f_4(2050) \to [\eta f_2(1270)]_{2,F}$ & & 1.1 &  0.5 \\
$\!\!\!\left. \begin{array}{l}
f_4(2050) \to [\omega\omega]_{2,D} \\
f_4(2050) \to [\omega\omega]_{2,G}
\end{array} \right\} $ &
$54\pm 13$ \tablenote{This number is the total for all partial waves.}&
$\begin{array}{c}
50 \\
2.0
\end{array} $ &
$\begin{array}{c}
10 \\
0.4
\end{array} $ \\
$f_4(2050) \to [K\bar{K}]_{0,G} $ & $1.4\pm 0.7$ & 5.4 & 1.6 \\
$f_4(2050) \to [K\bar{K}^*(892)+c.c.]_{1,G} $ & & 2.7 & 0.8 \\
$f_4(2050) \to [K \bar{K}_1(1270)+c.c.]_{1,F}$ & & 2.3  & 1.2 \\
$f_4(2050) \to [K^*(892) \bar{K}^*(892)]_{2,D}$ & & 7.3 & 2.1 \\
\tableline
$\sum_i\Gamma_i$ & $198 \pm 30$ & 435 & 109  
\end{tabular}
\end{table}

\newpage
\begin{table}
\caption{Calculated total decay widths (in MeV) for the $^3F_2$ and
$^3F_4$ $s\bar{s}$ states for different values of $\beta$. }
\label{table7}
\begin{tabular}{llll}
	$\beta$ (MeV) & $\gamma$ & $\Gamma (^3F_2)$ & $\Gamma (^3F_4)$ \\
\tableline
\multicolumn{4}{c}{RPSN} \\
\tableline
350 & 7.42 & 590 & 540 \\
400 & 9.73 & 1046 & 498 \\
450 & 12.0 & 1549 & 429 \\
481\tablenotemark[1] & 13.4 & 1841 & 388 \\
\tableline
\multicolumn{4}{c}{KIPSN} \\
\tableline
350 & 5.16 & 256 & 170 \\
371\tablenotemark[1] & 5.60 & 309 & 152 \\
400 & 6.25 & 391 & 132 \\
450 & 7.39 & 534 & 104 
\end{tabular}
\tablenotetext[1]{From the simultaneous fit of $\beta$ and $\gamma$.}
\end{table}

\end{document}